\documentclass[12pt]{article}
\usepackage[a4paper,top=2cm,bottom=2cm,left=3cm,right=1cm]{geometry}
\usepackage[pdftex]{graphicx}
\usepackage{amsmath}
\usepackage{amssymb}
\usepackage[usenames]{color}
\normalbaselines

\begin{document}
\newcommand{\pst}{\hspace*{1.5em}}
\newcommand{\be}{\begin{equation}}
\newcommand{\ee}{\end{equation}}
\newcommand{\bm}{\boldmath}
\newcommand{\ds}{\displaystyle}
\newcommand{\bdm}{\begin{displaymath}}
\newcommand{\edm}{\end{displaymath}}
\newcommand{\bea}{\begin{eqnarray}}
\newcommand{\eea}{\end{eqnarray}}
\newcommand{\ba}{\begin{array}}
\newcommand{\ea}{\end{array}}

\thispagestyle{plain}

\label{sh}

\begin{center} {\Large \bf
\begin{tabular}{c}
PROBABILITY REPRESENTATION OF QUANTUM 
\\[-1mm]
EVOLUTION AND ENERGY LEVEL EQUATIONS
\\[-1mm]
FOR OPTICAL TOMOGRAMS
\end{tabular}
 } \end{center}

\bigskip

\bigskip
\begin{center} {\bf
Ya. A. Korennoy, V.I.Man'ko
}\end{center}

\medskip

\begin{center}
{\it
P.N.    Lebedev Physics Institute,                          \\
       Leninsky prospect 53, 117924 Moscow, Russia
 }
\end{center}

\begin{abstract}\noindent
The von Neumann evolution equation for density matrix and the Moyal equation for the Wigner function are mapped onto evolution equation for optical tomogram of quantum state. The connection with known evolution equation for symplectic tomogram of the quantum state is clarified. The stationary states corresponding to quantum energy levels are associated with the probability representation of the von Neumann and Moyal equations written for the optical tomograms. Classical Liouville equation for optical tomogram is obtained. Example of parametric oscillator is considered in detail.
\end{abstract}


\medskip

\section{Introduction}
\pst
In \cite {Mancini 96}, \cite{Mancini Found Phys 97} new formulation of quantum mechanics called probability representation of quantum mechanics (see recent reviews \cite{Ibort Phys.Scr.}, \cite{Manko-Manko Found.Phys.2009}) was suggested. In the probability representation the quantum states are described by fair probability distributions called quantum tomograms or tomographic probability distributions. The tomograms contain complete information of the quantum state and they are connected with density operators of the states by means of invertable maps. There are several kinds of the tomograms related to the density operators by means of the different  invertable maps. The idea of introducing the probability representation of quantum mechanics was induced by existence of connection of the Wigner function $W(q,~p)$ of a system with continuous degrees of freedom like position with its Radom transform \cite{Radon17} $w(X,~\theta)$ called optical tomogram and found in \cite{Ber. Ber.}, \cite{Vog Ris}. The optical tomogram and its connection with the Wigner function was applied in experiments on homodyne detection of photon quantum states \cite{Raymer 93} (see review article \cite{Lvov. Ray Rev.Mod.Phys}), where the measured optical tomogram was considered as a technical tool to measure (reconstruct) the Wigner function of the photon state. 


In the approach based on reconstructing the Wigner function this quasidistribution function was considered as object identified with the quantum state and the optical tomogram was interpreted as intermediate step (not identified with the state) to obtain the Wigner function. In \cite{Mancini 95} the probability distribution called later \cite{d'Ariano} symplectic tomogram was introduced.  The symplectic tomogram $M(X,\mu,\nu)$ which is nonnegative probability distribution of homodine quadrature (of position) $X$ depending on extra real parameters $\mu$ and $\nu$ was shown to be related with optical tomogram and this relation provided the possibility to reconstruct the Wigner function using the symplectic tomogram by means of Fourier transform of the tomogram. 

Since symplectic and optical tomograms are connected by invertable maps with the Wigner function and, consequently, with the state density operator it was suggested to identified the quantum states with the tomographic probability distributions \cite{Mancini 96}, \cite{Mancini Found Phys 97},
\cite{Olga JRLR 97}, \cite{Mendes Physica}, \cite{Olga Tombesi JPA}.
It was shown that also in classical statistical mechanics the states can be described by the tomograms \cite{Olga JRLR 97}, \cite{Mendes Physica} connected with probability distributions $f(q,p)$ on the system phase-space by the Radon transform. Thus, it was understood that the concept of state in both classical and quantum mechanics is identified with the tomographic probability distribution. The basic equation of quantum mechanics, like quantum evolution equation for density matrix \cite{Von Neumann} and Moyal equation \cite{Moyal49} for the Wigner function \cite{Wigner32} were obtained in the probability representation for the symplectic tomograms \cite{Mancini 96}, \cite {Mancini Found Phys 97}, \cite{Olga JRLR 97}, \cite{Mendes Physica}, \cite{Ibort Phys.Scr.}.
Though the symplectic $M(X,\mu,\nu)$ and optical tomogram $w(X,\theta)$ are connected by the relation 
$w(X,\theta) = M(X,\cos\theta,\sin\theta)$; the evolution equation known for symplectic tomogram was not transformed in explicit form into the evolution equation for the optical tomogram.
Also the quantum equations for stationary states providing energy levels and written in the probability representation of quantum mechanics for symplectic tomograms (see, e.g. \cite{Ibort Phys.Scr.}) were not transformed till now into equation for the optical tomograms. On the other hand in experiments on homodyne detecting the photon states namely the optical tomogram is measured. In view of this one needs explicit form of the basic quantum equations written for the optical tomograms. Though, in principle, the equation written for symplectic tomogram implicitly contain information on the equation written for the optical tomogram there is a technical difficulty to obtain the equation since the optical tomogram depends on two variables and the symplectic tomogram depends on three variables. One needs to take into account this difference to do the calculations.
Thus the aim of our work is derive the evolution and energy level equations for the optical tomograms of quantum states for systems with several degrees of freedom and find the connection of these equations with equations for the symplectic tomograms.

The paper is organized as follows. In Sec.2 we review optical tomography of quantum states.
In Sec.3 and Sec.4 the connection of the Moyal equation with the quantum evolution for optical tomogram and symplectic tomogram is studied and energy level equation for optical tomogram is presented. In Sec.5 the classical Liouville equation for optical tomogram is derived. The tomograms of time dependent photon-added coherent states of parametric oscillator are found in Sec.6. The conclusion and prospects are presented in Sec.7.

\section{Symplectic and optical tomograms}
\pst
We review in this section the constructions of the optical and symplectic tomograms for system with one degree of freedom (or one-mode electromagnetic field). Given the state density operator $\hat\rho$ of the system. The symplectic tomogram $M(X,\mu,\nu)$ is defined as 
\be                             \label{eq_1}
M(X,\mu,\nu)=\langle\delta(X\hat1-\mu\hat q-\nu\hat p)\rangle \\
= \frac{1}{2\pi}\int{\mbox d}k~\mbox{Tr}\{\hat\rho~e^{ik(X\hat1-\mu\hat q-\nu\hat p)}\}.
\ee
Here $\hat q$ and $\hat p$ are position and momentum operators (photon quadratures), respectively, $\hat1$ is identity operator, and we assume that the taking trace and integration ower wave number $k$ are permutative operations for the states under consideration. The optical tomogram $w(X,\theta)$ is defined by analogous relation
\be                             \label{eq_2}
w(X,\theta)=\langle\delta(X\hat1-\hat q\cos\theta-\hat p\sin\theta)\rangle 
= \frac{1}{2\pi}\int{\mbox d}k~\mbox{Tr}\{\hat\rho~e^{ik(X\hat1-\hat q\cos\theta-\hat p\sin\theta)}\}.
\ee

The phase $0\leq\theta\leq2\pi$ is called in homodyne detecting the photon states as local oscillator phase and this parameter can be controlled and varied in experiments on measuring the quantum states \cite{Raymer 93},\cite{Lvov. Ray Rev.Mod.Phys}. Since the Dirac delta-function in Eq.(\ref{eq_1}) is homogeneous function,
i.e. $\delta(\lambda y)=|\lambda|^{-1}\delta(y)$ the symplectic tomogram is the homogeneous function
\be                             \label{eq_3}
M(\lambda X,\lambda\mu,\lambda\nu)=|\lambda |^{-1}M(X,\mu,\nu).
\ee
From the property of delta function $\int\delta(y-a){\mbox d}y=1$ it follows normalization condition for both symplectic tomogram and optical tomogram
\be                             \label{eq_4}
\int M(X,\mu,\nu)~{\mbox d}X=1
\ee
and
\be                             \label{eq_5}
\int w(X,\theta)~{\mbox d}X=1.
\ee
The homogenety condition (\ref{eq_3}) provides the relation of the optical and symplectic tomograms
\be                             \label{eq_6}
M(X,\mu,\nu)=(\mu^2+\nu^2)^{-1/2}~w\left(X(\mu^2+\nu^2)^{-1/2},\tan^{-1}\frac{\nu}{\mu}\right).
\ee
The Wigner function of state with density operator $\hat\rho$ is given by relation (we take $\hbar=1$)
\be                             \label{eq_7}
W(q,p)=\int \mbox{Tr}\left\{\hat\rho~\left|q-\frac{u}{2}\right\rangle\left\langle q+\frac{u}{2}\right|e^{-ipu}\right\}{\mbox d}u.
\ee
The Wigner function is normalized by the relation
\be		\label{eq_8}
\int~W(q,p)\frac{\mbox{d}q~\mbox{d}p}{2\pi}.
\ee
The symplectic tomogram is expressed in terms of the Wigner function in terms of its Radon transform depending on two parameters $\mu$ and $\nu$
\be		\label{eq_9}
M(X,\mu,\nu)=\int~W(q,p)\delta(X-\mu q-\nu p)\frac{\mbox{d}q~\mbox{d}p}{2\pi}.
\ee
The inverse transform is given by Fourier integral
\be		\label{eq_10}
W(q,p)=\frac{1}{2\pi}\int M(X,\mu,\nu)e^{i(X-\mu q-\nu p)}\mbox{d}X~\mbox{d}\mu~\mbox{d}\nu.
\ee
Analogously the optical tomogram is expressed in terms of Radon transform depending on the angle $\theta$
\be		\label{eq_11}
w(X,\theta)=\frac{1}{4\pi^2}\int W(q,p)e^{ik(X-q\cos\theta-p\sin\theta)}\mbox{d}q~\mbox{d}p~\mbox{d}k.
\ee
Using (\ref{eq_6}) one can to give the inverse Radon transform inserting the function $M(X,\mu,\nu)$ given by this relation into relation (\ref{eq_10}). Then one has
\be                             \label{eq_12}
W(q,p)=\frac{1}{2\pi}\int~(\mu^2+\nu^2)^{-1/2}~w\left(X(\mu^2+\nu^2)^{-1/2},\tan^{-1}\frac{\nu}{\mu}\right)
~e^{i(X-\mu q-\nu p)}~\mbox{d}X~\mbox{d}\mu~\mbox{d}\nu.
\ee
In this formula $-\infty< X,~\mu,~\nu~<\infty.$
One can transform the integral using polar coordinates
\be		\label{eq_13}
\mu = r\cos\theta,~~~\nu=r\sin\theta.
\ee
The symplectic tomogram is  even function. It means that the optical tomogram has the property
\be		\label{eq_14}
w(-X,\theta+\pi)=w(X,\theta).
\ee
Using this property the integral (\ref{eq_12}) can be presented in the form
\be		\label{eq_15}
W(q,p)=\frac{1}{2\pi}\int\limits_{0}^{\pi}~\mbox{d}\theta\int\limits_{-\infty}^{+\infty}\int\limits_{-\infty}^{+\infty}w(X,\theta)
|\eta |~e^{i\eta(X-q\cos\theta-p\sin\theta)} \mbox{d}\eta~\mbox{d}X.
\ee
The optical and symplectic tomograms are probability distributions of random quadrature $X$. Thus one has two characteristic functions for these distributions
\be		\label{eq_16}
\chi _M(z,\mu,\nu)=\int~M(X,\mu,\nu)~e^{izX}\mbox{d}X
\ee
and
\be		\label{eq_17}
\chi _w(\eta ,\theta)=\int~w(X,\theta)~e^{i\eta X}\mbox{d}X.
\ee
One has by using Fourier transforms the expressions 
\be		\label{eq_18}
M(X,\mu,\nu)=\frac{1}{2\pi}\int\chi_M(z,\mu,\nu)~e^{-izX}\mbox{d}z
\ee
and
\be		\label{eq_19}
w(X,\theta)=\frac{1}{2\pi}\int\chi_w(\eta ,\theta)~e^{-i\eta X}\mbox{d}\eta .
\ee
The characteristic functions are related
\be		\label{eq_20}
\chi_w(\eta,\theta)=\chi_M(\eta,\cos\theta,\sin\theta).
\ee
The highest momenta of the homodine quadrature $\langle X^n\rangle(\theta)$ are determined by the characteristic function $\chi_w(\eta,\theta)$, i.e.
\be		\label{eq_21}
\chi_w(\eta,\theta)=\sum\limits_{n=0}^{\infty}\frac{i^n}{n!}\eta^n\langle X^n\rangle(\theta).
\ee

\section{Evolution equation for tomograms}
\pst
In this section we derive the evolution and energy level equations for optical tomograms. The von Neumann evolution equation for density operator $\hat\rho(t)$ reads
\be		\label{eq_22}
\frac{\partial\hat\rho(t)}{\partial t}+i[\hat H,\hat\rho(t)]=0.
\ee
Here the Hamiltonian $\hat H=\hat p^2/2+U(\hat q)$ where $U(\hat q)$ is potential energy and the mass of the particle is taken as $m=1$.
In coordinate representation the von Neumann equation takes the form
\be		\label{eq_23}
i\frac{\partial\rho(x,x',t)}{\partial t}=-\frac{1}{2}\left(\frac{\partial^2}{\partial x^2}
-\frac{\partial^2}{\partial x'^2}\right)\rho(x,x',t)+\left(U(x)-U(x')\right)\rho(x,x',t).
\ee
Using the relation of the density matrix and the Wigner function 
\be		\label{eq_24}
\rho(x,x',t)=\frac{1}{2\pi}\int W\left(\frac{x+x'}{2},p,t\right)~e^{ip(x-x')}\mbox{d}p
\ee
one has the correspondence rules
\be		\label{eq_25}
\frac{\partial\rho}{\partial t} \longleftrightarrow\frac{\partial W}{\partial t},~~~~
\frac{\partial\rho}{\partial x} \longleftrightarrow\left(\frac{1}{2}\frac{\partial}{\partial q}+ip\right)W,
\ee
\bdm
\frac{\partial\rho}{\partial x'} \longleftrightarrow\left(\frac{1}{2}\frac{\partial}{\partial q}-ip\right)W;~~~~
x\rho \longleftrightarrow\left(q+\frac{i}{2}\frac{\partial}{\partial p}\right)W;~~~~
x'\rho \longleftrightarrow\left(q-\frac{i}{2}\frac{\partial}{\partial p}\right)W.
\edm
which can be used to get the von Neumann equation (\ref{eq_23}) in the Moyal form
\be		\label{eq_26}
\frac{\partial W(q,p,t)}{\partial t} + p\frac{\partial W(q,p,t)}{\partial q}
+\frac{1}{i}\left[U\left(q-\frac{i}{2}\frac{\partial}{\partial p}\right) - \mbox{c.c.}\right]W(q,p,t)=0.
\ee
Using the relation (\ref{eq_11}) one can find correspondence rules for operators acting on the Wigner function and the optical tomogram. In fact
\be		\label{eq_27}
\cos\theta\frac{\partial}{\partial X}w(X,\theta)=
\frac{1}{4\pi^2}\int ik\cos\theta~e^{-ikq\cos\theta}
W(q,p,t)~e^{ik(X-p\sin\theta)}\mbox{d}k~\mbox{d}q~\mbox{d}p.
\ee
In view of the equality 
\bdm
ik\cos\theta~e^{-ikq\cos\theta}=-\frac{\partial}{\partial q}~e^{-ikq\cos\theta}
\edm
and evaluating by parts the integral (\ref{eq_27}) over variables for functions 
$W(q,p)\to0$ for $q\to\pm \infty $ we get
\be		\label{eq_28}
\cos\theta\frac{\partial}{\partial X}w(X,\theta)=
\frac{1}{4\pi^2}\int \frac{\partial W(q,p)}{\partial q}\delta(X-q\cos\theta-p\sin\theta)\mbox{d}q~\mbox{d}p.
\ee
It means that
\be		\label{eq_29}
\frac{\partial}{\partial q}W(q,p) \longleftrightarrow\cos\theta\frac{\partial}{\partial X}w(X,\theta).
\ee
Analogously we have the correspondence rule
\be		\label{eq_30}
\frac{\partial}{\partial p}W(q,p) \longleftrightarrow\sin\theta\frac{\partial}{\partial X}w(X,\theta).
\ee
Applying the operator $\left(\partial /\partial X\right)^{-1}$ which is defind by acting on plane wave as 
\be		\label{eq_31}
\left(\frac{\partial}{\partial X}\right)^{-1}~e^{ikX}=\frac{1}{ik}~e^{ikX}
\ee
to (\ref{eq_11}) together with differentiation over the angle variable $\theta$ and multiplication by $\sin\theta$ we get the equality
\be		\label{eq_32}
\sin\theta\frac{\partial}{\partial \theta}\left(\frac{\partial}{\partial X}\right)^{-1}w(X,\theta)
+X\cos\theta~w(X,\theta)
=\frac{1}{4\pi^2}\int q~W(q,p)~e^{ik(X-q\cos\theta-p\sin\theta)}\mbox{d}k~\mbox{d}q~\mbox{d}p.
\ee
We used identities
\bdm
(q\sin\theta-p\cos\theta)\sin\theta+X\cos\theta=q+(X-q\cos\theta-p\sin\theta)\cos\theta
\edm
and
\bdm
\delta(X-q\cos\theta-p\sin\theta)~(X-q\cos\theta-p\sin\theta)=0.
\edm
The relation (\ref{eq_32}) gives the correspondence rule
\be		\label{eq_33}
q~W(q,p) \longleftrightarrow \left(\sin\theta\left(\frac{\partial}{\partial X}\right)^{-1}\frac{\partial}{\partial\theta}
+X\cos\theta\right)w(x,\theta).
\ee
Analogously we get the correspondence rule
\be		\label{eq_34}
p~W(q,p) \longleftrightarrow \left(-\cos\theta\left(\frac{\partial}{\partial X}\right)^{-1}\frac{\partial}{\partial\theta}
+X\sin\theta\right)w(x,\theta).
\ee
The correspondence rules (\ref{eq_29}), (\ref{eq_30}), (\ref{eq_33}), (\ref{eq_34})
give the possibility to transform the Moyal equation (\ref{eq_26}) into the evolution equation for the
optical tomogram $w(X,\theta)$. We get the new result in explicit form
\begin{eqnarray}                             
\frac{\partial}{\partial t}w(X,\theta,t)&=&
\left[\cos^2\theta\frac{\partial}{\partial\theta}
-\frac{1}{2}\sin2\theta\left\{1+X\frac{\partial}{\partial X}\right\}
\right]w(X,\theta,t) \nonumber \\[3mm]
&+&2\left[\mbox{Im}~U\left\{
\sin\theta\frac{\partial}{\partial\theta}
\left[\frac{\partial}{\partial X}\right]^{-1}
+X\cos\theta+i\frac{\sin\theta}{2}
\frac{\partial}{\partial X}\right\}\right]
w(X,\theta,t).\nonumber \\
		\label{eq_35}
\end{eqnarray}
This quantum evolution equation for the probability distribution $w(X,\theta,t)$ is compatible with the known evolution equation for the symplectic tomogram $M(X,\mu,\nu,t)$ which can be obtained from the Moyal equation in view of the correspondence rules
\begin{eqnarray}  
q~W(q,p) &\longleftrightarrow &-\left(\frac{\partial}{\partial X}\right)^{-1}\frac{\partial}{\partial\mu}M(X,\mu,\nu),
\nonumber \\[3mm]
p~W(q,p) &\longleftrightarrow &-\left(\frac{\partial}{\partial X}\right)^{-1}\frac{\partial}{\partial\nu}M(X,\mu,\nu),
\nonumber \\[3mm]
\frac{\partial}{\partial q}~W(q,p) &\longleftrightarrow &\mu\frac{\partial}{\partial X}M(X,\mu,\nu),
\nonumber \\[3mm]
\frac{\partial}{\partial p}~W(q,p) &\longleftrightarrow &\nu\frac{\partial}{\partial X}M(X,\mu,\nu).
		\label{eq_36}
\end{eqnarray}
These rules are easily obtained by considering Eq.(\ref{eq_9}) where delta function is presented by the Fourier integral, i.e.
\be		\label{eq_37}
M(X,\mu,\nu)=\frac{1}{4\pi^2}\int~W(q,p)~e^{ik(X-\mu q-\nu p)}\mbox{d}k~\mbox{d}q~\mbox{d}p.
\ee
For example the application to both sides of Eq.(\ref{eq_37}) the operator 
$\frac{\partial}{\partial\mu}\left(\frac{\partial}{\partial X}\right)^{-1}$
we have
\be		\label{eq_38}
\frac{\partial}{\partial\mu}\left(\frac{\partial}{\partial X}\right)^{-1}M(X,\mu,\nu)=
\frac{1}{4\pi^2}\int(-q)~W(q,p)~e^{ik(X-\mu q-\nu p)}\mbox{d}k~\mbox{d}q~\mbox{d}p,
\ee
and equality gives first correspondence rule in (\ref{eq_36}). Analogously we get other correspondence rules 
in (\ref{eq_36}). Applying the correspondence rules to Moyal evolution equation we get the evolution equation for the symplectic tomogram $M(X,\mu,\nu)$, i.e. 
\begin{eqnarray}
\frac{\partial}{\partial t}M(X,\mu,\nu,t)&=&
\mu\frac{\partial}{\partial\nu}
M(X,\mu,\nu,t) \nonumber \\[3mm]
&+&
2\left[\mbox{Im}~
U\left\{-\left[\frac{\partial}{\partial X}\right]^{-1}
\frac{\partial}{\partial\mu}+\frac{i\nu}{2}
\frac{\partial}{\partial X}\right\}\right]
M(X,\mu,\nu,t).
		\label{eq_39}
\end{eqnarray}

Using connection of symplectic and optical tomograms (\ref{eq_6}) one can derive 
(\ref{eq_35}) from (\ref{eq_39}).
The equation (\ref{eq_35}) from (\ref{eq_39}) can be transformed into equations for characteristic functions 
(\ref{eq_16}) and (\ref{eq_17}).
This can be done by substitutions
\bdm
\frac{\partial}{\partial X} \rightarrow iz,~~~~\left(\frac{\partial}{\partial X}\right)^{-1} \rightarrow \frac{1}{iz}
\edm
in (\ref{eq_39}) in the case of the function $\chi _M(z,\mu,\nu)$ and
\bdm
\frac{\partial}{\partial X} \rightarrow i\eta,~~~~\left(\frac{\partial}{\partial X}\right)^{-1} \rightarrow \frac{1}{i\eta},~~~~
X \rightarrow i\frac{\partial}{\partial \eta} .
\edm
Thus we get the evolution equation
\begin{eqnarray}           	
\frac{\partial\chi _M(z,\mu,\nu,t)}{\partial t}&=&
\nu\frac{\partial}{\partial\mu}\chi_M(z,\mu,\nu,t) \nonumber \\[3mm]
&-&\frac{1}{i}\left[U\left\{-\frac{1}{iz}\frac{\partial}{\partial\mu}+\frac{\nu z}{2}\right\}
- U\left\{-\frac{1}{iz}\frac{\partial}{\partial\mu}-
\frac{\nu z}{2}\right\}\right]
\chi_M(z,\mu,\nu,t)
		\label{eq_40}
\end{eqnarray}
for the function $\chi _M(z,\mu,\nu,t)$. For the optical tomographic characteristic function we obtain the evolution equation
\begin{eqnarray}           	
\frac{\partial\chi _w(\eta,\theta,t)}{\partial t}&=&
\left[\cos^2\theta\frac{\partial}{\partial\theta}
-\sin2\theta\left(1+\frac{\eta}{2}\frac{\partial}{\partial\eta}\right)\right]\chi_w(\eta,\theta,t) \nonumber \\[3mm]
&-&\frac{1}{i}\left[U\left\{\frac{\sin\theta}{i\eta}\frac{\partial}{\partial\theta}
+i\cos\theta\frac{\partial}{\partial\theta}+\frac{\sin\theta}{2}\eta\right\}\right.\nonumber \\[3mm]
&&\left.- U\left\{\frac{\sin\theta}{i\eta}\frac{\partial}{\partial\theta}
+i\cos\theta\frac{\partial}{\partial\theta}-\frac{\sin\theta}{2}\eta\right\}\right]
\chi_w(\eta,\theta,t).
		\label{eq_41}
\end{eqnarray}

\section{Multidimensional case}
\pst
The results of previous sections  can be generalized to multidimensional case.
Thus, for multimode  Hamiltonians
\be		\label{eq_42}
\hat H=\sum_{\sigma=1}^{n}\frac{\hat p_\sigma^2}{2m_\sigma}~+~U(\vec q),~~~~\vec q=\{q_\sigma\},
\ee
and for optical and symplectic multidimensional tomograms
\be		\label{eq_43}
w(\vec X,\vec\theta,t)=\int~W(\vec q,\vec p,t)\prod_{\sigma=1}^{n}
\exp\left\{ik_\sigma\left(X_\sigma-q_\sigma\cos\theta_\sigma
-p_\sigma\frac{\sin\theta_\sigma}{m_\sigma\omega_\sigma}\right)\right\}
\frac{\mbox{d}^nk~\mbox{d}^nq~\mbox{d}^np}{(2\pi\hbar)^{2n}};
\ee
\be		\label{eq_44}
M(\vec X,\vec\mu,\vec\nu,t)=\int~W(\vec q,\vec p,t)\prod_{\sigma=1}^{n}
\exp\{ik_\sigma(X_\sigma-\mu_\sigma q_\sigma-\nu_\sigma p_\sigma)\}\frac{\mbox{d}^nk~\mbox{d}^nq~\mbox{d}^np}{(2\pi\hbar)^{2n}},
\ee
where $\omega_{\sigma}$ and $m_\sigma$ are frequency and mass dimensional constants respectively
for $\sigma-$th degree of freedom ($\omega_{\sigma}$ are choosing from convenience considerations for particular Hamiltonian),
multidimensional analogs of the equations (\ref{eq_35}) and (\ref{eq_39}) can be written in the forms:
\begin{eqnarray}                             
\frac{\partial}{\partial t}w({\vec{X}},{\vec\theta},t)&=&
\left[\sum_{\sigma=1}^n\omega_{\sigma}
\left[\cos^2\theta_\sigma\frac{\partial}{\partial\theta_\sigma}
-\frac{1}{2}\sin2\theta_\sigma\left\{1+X_\sigma\sin\theta_\sigma\frac{\partial}{\partial X_\sigma}\right\}
\right]\right]w({\vec X},{\vec\theta},t)\nonumber \\[3mm]
&+&\frac{2}{\hbar}\left[\mbox{Im}~U\left\{
\sin\theta_\sigma\frac{\partial}{\partial\theta_\sigma}
\left[\frac{\partial}{\partial X_\sigma}\right]^{-1}
+X_\sigma\cos\theta_\sigma+i\frac{\hbar\sin\theta_\sigma}
{2m_\sigma\omega_{\sigma}}
\frac{\partial}{\partial X_\sigma}\right\}\right]
w({\vec X},{\vec\theta},t);\nonumber \\
\label{eq_45}
\end{eqnarray}
\begin{eqnarray}
\frac{\partial}{\partial t}M({\vec X},{\vec\mu},{\vec\nu},t)&=&
\left[\sum_{\sigma=1}^n\frac{\mu_\sigma}{m_\sigma}\frac{\partial}{\partial\nu_\sigma}\right]
M({\vec X},{\vec\mu},{\vec\nu},t) \nonumber \\[3mm]
&+&
\frac{2}{\hbar}
\left[\mbox{Im}~
U\left\{-\left[\frac{\partial}{\partial X_\sigma}\right]^{-1}
\frac{\partial}{\partial\mu_\sigma}+\frac{i\nu_\sigma\hbar}{2}
\frac{\partial}{\partial X_\sigma}\right\}\right]
M({\vec X},{\vec\mu},{\vec\nu},t).
\label{eq_46}
\end{eqnarray}
We reconstructed in these formulas dimensional constants including the Planck constant. Note, that the
parameters $\mu_\sigma$ are dimensionless, while parameters $\nu_\sigma$ have dimension of $[m]^{-1}[t]$.

For completeness we present also energy level equations for multimode Hamiltonians (\ref{eq_42}),
when $U(\vec q)$ is independent of time.
The equations have the form:
\begin{eqnarray}
E~w_E({\vec X},{\vec\theta})&=&\left[\sum_{\sigma=1}^nm_\sigma\omega_\sigma^2\left\{\frac{\cos^2\theta_\sigma}{2}
\left[\frac{\partial}{\partial X_\sigma}\right]^{-2}
\left(\frac{\partial^2}{\partial\theta_\sigma^2}+1\right)\right.\right.\nonumber \\[3mm]
&-&\left.\frac{X_\sigma}{2}\left.\left[\frac{\partial}{\partial X_\sigma}\right]^{-1}
\left(\cos^2\theta_\sigma+\sin2\theta_\sigma
\frac{\partial}{\partial\theta_\sigma}\right)
+\frac{X_\sigma^2}{2}\sin^2\theta_\sigma-\frac{1}{8}\cos^2\theta_\sigma
\frac{\partial^2}{\partial X^2_\sigma}\right\}\right]
w_E({\vec X},{\vec\theta}) \nonumber \\[3mm]
&+&\left[\mbox{Re}~U\left\{\sin\theta_\sigma
\frac{\partial}{\partial\theta_\sigma}
\left[\frac{\partial}{\partial X_\sigma}\right]^{-1}
+X_\sigma\cos\theta_\sigma+i\frac{\hbar\sin\theta_\sigma}{2m_\sigma\omega_\sigma}
\frac{\partial}{\partial X_\sigma}\right\}\right]
w_E({\vec X},{\vec\theta});
\label{eq_47}
\end{eqnarray}
\begin{eqnarray}
E~M_E(\vec X,\vec\mu,\vec\nu)&=&
\left[\sum_{\sigma=1}^n \left\{\frac{1}{2m_\sigma}\left[\frac{\partial}{\partial X_\sigma}\right]^{-2}\left[\frac{\partial}{\partial\nu_\sigma}\right]^2-
\frac{\mu_\sigma^2\hbar^2}{8m_\sigma}\left[\frac{\partial}{\partial X_\sigma}\right]^2\right\}\right] M_E(\vec X,\vec\mu,\vec\nu) \nonumber \\[3mm]
&+&\left[{\mbox{Re}}~U\left\{
-\left[\frac{\partial}{\partial X_\sigma}\right]^{-1}\frac{\partial}{\partial\mu_\sigma}+
\frac{i\nu_\sigma\hbar}{2}\frac{\partial}{\partial X_\sigma}
\right\}\right] M_E(\vec X,\vec\mu,\vec\nu).
\label{eq_48}
\end{eqnarray}
The stationarity conditions of the distributions $w_E({\vec X},{\vec\theta})$ and $M_E(\vec X,\vec\mu,\vec\nu)$
are obtained from the equations 
(\ref{eq_45}) and (\ref{eq_46}), under conditions $\partial_tw_E=0$ and $\partial_tM_E=0$ respectively.
Thus each tomogram of stationary state of quantum system obey two equations simultaneously:
energy level equation and stationarity condition.

\section{Classical Liouville equation in tomographic form}
\pst
Classical Liouville equation in phase space for the  potential $U(\vec q)$ has the form
\be		\label{eq_49}
\frac{\partial f(\vec q,\vec p,t)}{\partial t} + \sum_{\sigma=1}^n \frac{p_\sigma}{m_\sigma}
\frac{\partial f(\vec q,\vec p,t)}{\partial q_\sigma}
-\sum_{\sigma=1}^n  \frac{\partial U(\vec q,t)}{\partial q_\sigma}\frac{\partial f(\vec q,\vec p,t)}{\partial p_\sigma}=0,
\ee
where $n$ - is a number of degrees of freedom.
If we introduce optical and symplectic tomograms of the distribution function $f$ as follows:
\be		\label{eq_50}
w_{cl}(\vec X,\vec\theta,t)=\frac{1}{(2\pi)^n}\int f(\vec q,\vec p,t)
\prod_{\sigma=1}^{n} \exp\left\{ik_\sigma\left(X_\sigma
-q_\sigma\cos\theta_\sigma-p_\sigma\frac{\sin\theta_\sigma}{m_\sigma\omega_\sigma}\right)\right\}
\mbox{d}^nq~\mbox{d}^np~\mbox{d}^nk;
\ee
\be		\label{eq_51}
M_{cl}(\vec X,\vec\mu,\vec\nu,t)=\frac{1}{(2\pi)^n}\int~f(\vec q,\vec p,t)
\prod_{\sigma=1}^{n}\exp\left\{ik_\sigma\left(X_\sigma-\mu_\sigma q_\sigma
-\nu_\sigma p_\sigma\right)\right\}\mbox{d}^nk~\mbox{d}^nq~\mbox{d}^np,
\ee
the inverse transforms are given by the Fourier integrals: 
\be		\label{eq_52}
f(\vec q,\vec p,t)=\int\limits_{0}^{\pi}\mbox{d}^n\theta
\int\limits_{-\infty}^{+\infty}\frac{\mbox{d}^n\eta~\mbox{d}^nX}{(2\pi)^{2n}}
w_{cl}(\vec X,\vec\theta,t)\prod_{\sigma=1}^{n}
\frac{|\eta_\sigma|}{m_\sigma\omega_\sigma}\exp\left\{ik_\sigma\left(X_\sigma
-q_\sigma\cos\theta_\sigma-p_\sigma\frac{\sin\theta_\sigma}{m_\sigma\omega_\sigma}\right)\right\} ;
\ee
\be		\label{eq_53}
f(\vec q,\vec p,t)=\frac{1}{(2\pi)^{2n}}\int M_{cl}(\vec X,\vec\mu,\vec\nu,t)
\prod_{\sigma=1}^{n}\exp\left\{ik_\sigma\left(X_\sigma-\mu_\sigma q_\sigma
-\nu_\sigma p_\sigma\right)\right\}\mbox{d}^nX~\mbox{d}^n\mu~\mbox{d}^n\nu.
\ee
As the relations (\ref{eq_50}) and (\ref{eq_51}) for $f(\vec q,\vec p)$
are similar to the relations (\ref{eq_43}) and (\ref{eq_44})
for the Wigner function $W(\vec q,\vec p)$ we can use multimode analogs of the correspondence rules  (\ref{eq_29}), (\ref{eq_30}), (\ref{eq_33}), (\ref{eq_34}) and (\ref{eq_36}) to obtain evolution equations for the tomograms $w_{cl}$ and $W_{cl}$ from the Liouville equation (\ref{eq_42}). After calculations we get:
\begin{eqnarray}                             
\frac{\partial}{\partial t}w_{cl}(\vec X,\vec\theta,t)&=&
\sum_{\sigma=1}^{n}\omega_\sigma \left[\cos^2\theta_\sigma\frac{\partial}{\partial\theta_\sigma}
-\frac{1}{2}\sin2\theta_\sigma\left\{1+X_\sigma\frac{\partial}{\partial X_\sigma}\right\}
\right]w_{cl}(\vec X,\vec\theta,t) \nonumber \\[3mm]
&+&\left[\sum_{\sigma=1}^{n}\frac{\partial}{\partial q_\sigma}~U\left\{
q_\sigma\rightarrow\sin\theta_\sigma\frac{\partial}{\partial\theta_\sigma}
\left[\frac{\partial}{\partial X_\sigma}\right]^{-1}
+X_\sigma\cos\theta_\sigma\right\}
\frac{\sin\theta_\sigma}{m_\sigma\omega_\sigma}\frac{\partial}{\partial X_\sigma}\right]
w_{cl}(\vec X,\vec\theta,t);\nonumber \\
		\label{eq_54}
\end{eqnarray}
\begin{eqnarray}
\frac{\partial}{\partial t}M_{cl}(\vec X,\vec\mu,\vec\nu,t)&=&
\vec\mu\frac{\partial}{\partial\vec\nu}
M_{cl}(\vec X,\vec\mu,\vec\nu,t) \nonumber \\[3mm]
&+&
\left[\sum_{\sigma=1}^{n}\frac{\partial}{\partial q_\sigma}U\left\{q_\sigma\rightarrow-\left[\frac{\partial}{\partial X_\sigma}\right]^{-1}
\frac{\partial}{\partial\mu_\sigma}\right\}
\nu_\sigma\frac{\partial}{\partial X_\sigma}\right]
M_{cl}(\vec X,\vec\mu,\vec\nu,t).
		\label{eq_55}
\end{eqnarray}
Note, that these equations are the limit cases of the corresponding equations
(\ref{eq_45}) and (\ref{eq_46}) when Planck constant is taken to be zero. Other classical kinetic equations can be presented in the analogous tomographic form.

\section{Tomograms of time dependent photon-added coherent states of parametric oscillator}

As an example let us find tomograms of studied in \cite{Phys.Scr.phot.ad.} time dependent photon-added
coherent states $|\alpha,m,t\rangle$ of one mode parametric oscillator with the Hamiltonian
\bdm
\hat H=\frac{\hat p^2}{2}+~\Omega^2(t)\frac{\hat q^2}{2},~~~~\Omega(0)=1.
\edm
The state $|\alpha,m,t\rangle$ is defined as follows:
\be				\label{Ua}
\vert\alpha,m,t\rangle=~\hat U(t)\vert\alpha,m,0\rangle= 
~\hat U(t)\vert\alpha,m\rangle=
\left(m!L_m(-|\alpha|^2)\right)^{-1/2}\hat U(t)
\hat a^{+m}\vert\alpha\rangle,
\ee
where $L_m(z)\equiv L_m^{(0)}(z)$ is the Laguerre polynomial \cite{Bateman,Szego},
$\vert\alpha\rangle$ is the initial coherent state,  $\hat U(t)$ is the unitary evolution operator
\be				\label{UU+}
~\hat U(t)~\hat U^+(t)=~\hat 1, 
~~~~~~\hat U(0)=~\hat 1.
\ee
As shown in \cite{Phys.Scr.phot.ad.}  the  expression for
the state  $\vert\alpha,m,t\rangle$ in the coordinate representation has the form
\be
\langle q\vert\alpha,m,t\rangle=
\left(m!L_m(-|\alpha|^2)\right)^{-1/2}
\left(\ds{\frac{\varepsilon^*}{2\varepsilon}}\right)^{m/2}
~H_m\left(\ds{\frac{q}{\vert\varepsilon\vert}-\sqrt{\frac{\varepsilon^*}
{2\varepsilon}}\alpha}\right)\langle q\vert\alpha,t\rangle,
\ee
where $H_m(z)$ is the Hermite polynomial \cite{Bateman,Szego}, $\langle q\vert\alpha,t\rangle$ is the time-dependent coherent state
\be				\label{timeCohSt}
\langle q\vert\alpha,t\rangle =
\pi^{-1/4}\varepsilon ^{-1/2}
\exp\left(\frac{i\dot\varepsilon q^2}{2\varepsilon}+
\frac{\sqrt 2\alpha q}{\varepsilon}
-\frac{\alpha^2\varepsilon^*}{2\varepsilon}
-\frac{\vert\alpha\vert^2}{2}\right), 
\ee
which was found for the first time in \cite{MM70},
 $c$-number function $\varepsilon(t)$ satisfies the equation
\be			\label{e-equation}
~\ddot\varepsilon(t)+~\Omega^2(t)~\varepsilon(t)=0,
\ee
with the initial conditions $~\varepsilon(0)=1, ~~~\dot\varepsilon(0)=i,~~$
which means that the Wronskian is
\be				\label{Vronscian}
\varepsilon\dot\varepsilon^*-
\varepsilon^*\dot\varepsilon=-2i.
\ee
The state $|\alpha,m,t\rangle$ is pure, thus we can find tomogram of it from the formular
\bdm
M(X,\mu,\nu,t)=\frac{1}{2\pi|\nu|}\left|\int\exp\left(
\frac{i}{\nu}qX-\frac{i\mu}{2\nu}q^2\right)
\langle q\vert\alpha,m,t\rangle~{\mbox d}q\right|^2.
\edm
After some calculations we get
\begin{eqnarray}
M_{\alpha m}(X,\mu,\nu,t)&=&\frac{\left(m!L_m(-|\alpha|^2)\right)^{-1}}{\sqrt\pi\sqrt{2^m}
|\dot\varepsilon\nu+\varepsilon\mu|}
\left|H_m\left\{\left(\frac{X\varepsilon+i\sqrt2\alpha\nu}
{|\varepsilon|(\mu\varepsilon+\nu\dot\varepsilon)}
-\sqrt{\frac{\varepsilon^*}{2\varepsilon}}\alpha\right)
\left(\frac{|\varepsilon|^2(\mu\varepsilon+\nu\dot\varepsilon)}
{\varepsilon^2(\mu\varepsilon^*+\nu\dot\varepsilon^*)}\right)^{1/2}
\right\}\right|^2 \nonumber \\[3mm]
{}&\otimes&\left|\exp\left\{-\frac{|\alpha|^2}{2}
-\frac{X^2}{2|\mu\varepsilon+\nu\dot\varepsilon|^2}
+\frac{\sqrt2\alpha X}{\mu\varepsilon+\nu\dot\varepsilon}
-\frac{\alpha^2\varepsilon^*}{2\varepsilon}
+\frac{i\nu\alpha^2}{\varepsilon(\mu\varepsilon+\nu\dot\varepsilon)}
\right\}\right|^2.
\label{eq_62}
\end{eqnarray}
The substitutions $\mu=\cos\theta$ and $\nu=\sin\theta$ to (\ref{eq_62})
gives us the optical tomogram $w_{\alpha m}(X,\theta,t).$

\section{Conclusion}
\pst
To summarize, we point out the main results of this work.

We obtained the evolution equations for optical tomogram of state of quantum and classical systems both one dimensional and multidimensional cases. We shown the correspondence between the evolution equations for optical and symplectic tomograms.
The importance of the quantum tomographic equations written namely for optical tomogram is connected
with the fact that namely these tomograms are measured in experiments to get the photon states and to study their characteristics. The evolution equation found in the work provide the possibility of monitoring the system quantum
states in the process of the time evolution. The generalization of the results to relativistic like in \cite{Chernega},
\cite{Arhopov} kinetic equations will be given in further publications.


\begin{thebibliography}{99}

\bibitem{Mancini 96} S. Mancini, V. I. Man'ko and P. Tombesi,  \textsl{Phys.~Lett.}~A, \textbf{213}:1-2 1--6 (1996).

\bibitem{Mancini Found Phys 97} S. Mancini, V. I. Man'ko, and P. Tombesi, \textsl{Found.~Phys.}, \textbf{27}
801--824 (1997).

\bibitem{Ibort Phys.Scr.} A. Ibort, V. I. Man'ko, G. Marmo, A. Simoni, and F. Ventriglia, \textsl{Phys.~Scr.}, \textbf{79}, 065013 (2009).

\bibitem{Manko-Manko Found.Phys.2009} M. A. Man'ko and V. I. Man'ko,  \textsl{Found.~Phys.}, doi:10.1007/s10701-009-9403-9 (2009).

\bibitem{Radon17} J. Radon, \textsl{Ber. Verh. Sachs. Akad.}, \textbf{69}, 262 (1917).

\bibitem{Ber. Ber.}J. Bertrand and P. Bertrand, \textsl{Found. Phys.}, \textbf{17}, 397 (1987) .

\bibitem{Vog Ris} K. Vogel and H. Risken, Phys. Rev. A \textbf{40}, 2847 (1989) .

\bibitem{Raymer 93} D. T. Smithey, M. Beck, M. G. Raymer, A. Faridani, \textsl{Phys. Rev. Lett.}, \textbf{70},
1244 (1993).

\bibitem{Lvov. Ray Rev.Mod.Phys} A. I. Lvovsky and M. G. Raymer, \textsl{Rev. Mod. Phys.}, \textbf{81}, 299 (2009).

\bibitem{Mancini 95} S. Mancini, V. I. Man'ko and P. Tombesi, \textsl{Quantum~Semiclass.~Opt.},
\textbf{7}, 615 (1995). 

\bibitem{d'Ariano} G. M. D'Ariano, S. Mancini, V. I. Man'ko and P. Tombesi, \textsl{J. Opt. B:
Quantum Semiclass. Opt.}, \textbf{8}, 1017 (1996).

\bibitem{Olga JRLR 97} O. V. Man'ko and V. I. Man'ko, \textsl{J. Russ. Laser Res.}, \textbf{18}, 407 (1997).

\bibitem{Mendes Physica} V. I. Man'ko and R. V. Mendes, \textsl{Physica D}, \textbf{145}, 330 (2000).

\bibitem{Olga Tombesi JPA} S. Mancini, O. V. Man'ko, V. I. Man'ko, P. Tombesi, \textsl{J. Phys. A: Math. Gen.}, \textbf{34}:16 3461-3476 (2001).

\bibitem{Von Neumann} J. von Neumann, {\it Mathematische Grundlagen der Quantenmechanik},
Springer, Berlin (1932).

\bibitem{Moyal49} J. E. Moyal, \textsl{Proc. Cambrege Philos. Soc.}, \textbf{45}, 99 (1949).

\bibitem{Wigner32} E. Wigner,   \textsl{Phys. Rev.}, \textbf{40}, 749, (1932).

\bibitem{Chernega},V. N. Chernega, V. I. Man'ko \textsl{J. Russ. Laser Res.}, \textbf{29}, 43 (2008).

\bibitem{Arhopov} A. S. Arkhopov, V. I. Man'ko, \textsl{J. Russ. Laser Res.}, \textbf{25}, 468 (2004).

\bibitem{Phys.Scr.phot.ad.} V. V. Dodonov, M. A. Marchiolli, Ya. A. Korennoy,
V.I. Man'ko, E. A. Moukhin, \textsl{Phys.~Scr.}, \textbf{58}, 469 (1998).

\bibitem{Bateman} {\it Bateman Manuscript Project: Higher
Transcendental Functions\/}, edited by A.~Erd\'elyi
(McGraw-Hill, New York, 1953).

\bibitem{Szego} G.~Szeg\"o, {\it Orthogonal Polynomials\/} (American 
Mathematical Society, Providence, RI, 1959).

\bibitem{MM70} I.A.Malkin and V.I.Man'ko, Phys. Lett. A {\bf 31}, 243 (1970).



\end{thebibliography}
\end{document}